\documentclass[12pt,twoside]{article}
\usepackage{xrb2007}
\begin{document}

\session{Faint XRBs and Galactic LMXBs}

\shortauthor{Wachter}
\shorttitle{Spitzer Observations of LMXBs}

\title{Spitzer Space Telescope Observations of Low Mass X-ray Binaries}
\author{Stefanie Wachter}
\affil{Spitzer Science Center, California Institute of Technology, 1200 E. California Blvd. Pasadena, CA 91125}

\vspace{0.7cm}

\begin{abstract}
We present preliminary results from our archival Spitzer Space Telescope 
program aimed at characterizing the mid-IR properties of compact objects, 
both isolated and in binary systems, i.e. white dwarfs, X-ray binaries, 
cataclysmic variables, and magnetars. Most of these sources are too faint 
at mid-IR wavelengths to be observable from the ground, so this study 
provides the very first comprehensive look at the mid-IR emission of these 
objects. Here we present our results for the low mass X-ray binaries (LMXBs). 
We considered all of the systems listed in the most recent catalog of Liu 
et al. (2007) that have known optical counterparts. The particular goals 
of our projects encompass:  to establish the mid-IR spectral energy 
distribution, to search for the signatures of jets, circumbinary disks, 
low mass or planetary companions and debris disks, and to study the local 
environment of these sources.

\end{abstract}

\vspace{-0.1cm}

\section{Introduction}

We have searched the Spitzer Space Telescope archive (with {\em Leopard}) for observations
covering the positions of all low mass X-ray binaries with known optical counterparts listed
in Liu et al. (2007). Table 1 details the data available for each source 
as of March 21, 2008. Note that not all of the data listed in the table are public at the time
of this writing. Accepted Cycle 5 programs are listed, but not yet observed.
In Table 2 we provide results as to whether a particular 
source is detected or not, together with published Spitzer measurements of LMXBs.  
In this context, (f) denotes faint, in a most 
subjective fashion. (s) indicates that some pixels of the source PSF are saturated. 
We have found 17 newly detected sources in addition to the eight published Spitzer detections of LMXBs. 
Photometry of these sources is still ongoing and will be published at a later date.

\vspace{0.8cm}

\acknowledgements The research described in this poster  was carried out, in part, at the 
Jet Propulsion Laboratory, California Institute of Technology, and was sponsored
by the National Aeronautics and Space Administration through a grant to the PI. We made use 
of data products from the 2 Micron All Sky Survey, which is a joint project of the University
of Massachusetts and the Infrared Processing and Analysis Center/California Institute of Technology, 
funded by the National Aeronautics and Space
Administration and the National Science Foundation. 
It also utilized NASA's Astrophysics Data System Abstract Service and the SIMBAD database operated by
CDS, Strasbourg, France.

\begin{table}
\caption{LMXBs observed by the Spitzer Space Telescope}
\scriptsize
\begin{tabular}{lcrrl}
                &              &             &               & \\    
{\bf Name\tablenotemark{a}} & {\bf Altern. Name}& {\bf RA (2000)} & {\bf DEC (2000)} & {\bf PID/AORkey, Instr. Mode, Notes} \\ \hline
                &              &             &               & \\
IGR J00291+5934 &              & 00 29 03.06 & 59 34 19.00   &  No programs found\\
4U 0042+32      &              & 00 44 50.40 & 33 01 17.00   &  No programs found\\
GRO J0422+32    &              & 04 21 42.79 & 32 54 27.10   &  40948/23365888  IRAC Mapping\\
2A 0521$-$720   & LMC X$-$2    & 05 20 28.13 &$-$71 57 35.34 &  20203/14358272  IRAC Mapping\\ 
                &              &             &               &  20203/14370816  IRAC Mapping\\
                &              &             &               &  20203/14356480  IRAC Mapping\\
                &              &             &               &  20203/14371072  IRAC Mapping \\
                &              &             &               &  20203/14390016  MIPS Scan Map\\
                &              &             &               &  20203/14383104  MIPS Scan Map\\
                &              &             &               &  20203/14383360  MIPS Scan Map\\
                &              &             &               &  20203/14389760  MIPS Scan Map\\
4U 0614+091     &              & 06 17 07.40 & 09 08 13.60   &  20224/14415872  IRAC Mapping \\
                &              &             &               &  30803/19018240  IRAC Mapping\\
                &              &             &               &  30803/19018496  MIPS Photometry\\
1A 0620$-$00    &              & 06 22 44.50 &$-$00 20 44.72 &  3289/10702592  MIPS Photometry\\
                &              &             &               &  3289/10703616  IRAC Mapping\\
                &              &             &               &  30616/18618112  IRAC Mapping\\
                &              &             &               &  30616/18618624  IRAC Mapping\\
EXO 0748$-$676  &              & 07 48 33.80 &$-$67 45 08.93 &  No programs found\\
4U 0919$-$54    &              & 09 20 26.95 &$-$55 12 24.70 &  20224/14416128  IRAC Mapping\\ 
2S 0921$-$630   &              & 09 22 34.73 &$-$63 17 41.37 &  No programs found\\
XTE J0929$-$314 &              & 09 29 20.19 &$-$31 23 03.20 &  No programs found\\
GRS 1009$-$45   &              & 10 13 36.30 &$-$45 04 32.00 &  40948/23364864  IRAC Mapping\\
XTE J1118+480   &              & 11 18 10.85 & 48 02 12.90   &  3289/10702848  MIPS Photometry\\
                &              &             &               &  3289/10703872  IRAC Mapping\\ 
GS 1124$-$684   &              & 11 26 26.70 &$-$68 40 32.60 &  40948/23365376  IRAC Mapping\\
1A 1246$-$588   &              & 12 49 39.36 &$-$59 05 14.68 &  No programs found\\
4U 1254$-$69    &              & 12 57 37.20 &$-$69 17 20.80 &  No programs found\\
4U 1323$-$62    &              & 13 26 36.31 &$-$62 08 09.90 &  190/9231616  IRAC Mapping\\
                &              &             &               &  190/9234176  IRAC Mapping\\
                &              &             &               &  20597/15646208  MIPS Scan Map\\
                &              &             &               &  20597/15657216  MIPS Scan Map\\
GS 1354$-$64    &              & 13 58 09.92 &$-$64 44 04.90 &  No programs found\\ 
4U 1456$-$32    &  Cen X$-$4   & 14 58 22.00 &$-$31 40 08.00 &  3289/10703360  MIPS Photometry\\
                &              &             &               &  3289/10704384  IRAC Mapping\\
3A 1516$-$569   &  Cir X$-$1   & 15 20 40.90 &$-$57 10 01.00 &  190/9226752  IRAC Mapping\\ 
                &              &             &               &  190/9228032  IRAC Mapping\\
                &              &             &               &  191/9331200  IRAC Mapping\\
                &              &             &               &  20597/15593984  MIPS Scan Map\\
                &              &             &               &  20597/15626240  MIPS Scan Map\\
1A 1524$-$61    &  TrA X$-$1   & 15 28 17.20 &$-$61 52 58.00 &  50477/26257920  MIPS Scan Map\\
4U 1543$-$47    &              & 15 47 08.60 &$-$47 40 10.00 &  No programs found\\
4U 1543$-$624   &              & 15 47 54.29 &$-$62 34 05.13 &  20224/14415616  IRAC Mapping\\
XTE J1550$-$564 &              & 15 50 58.78 &$-$56 28 35.00 &  No programs found\\
4U 1556$-$60    &              & 16 01 02.30 &$-$60 44 18.00 &  No programs found\\
1E 1603.6+2600  &              & 16 05 45.82 &   25 51 45.10 &  No programs found\\
4U 1608$-$52    &              & 16 12 43.00 &$-$52 25 23.00 &  191/9329664  IRAC Mapping\\
                &              &             &               &  191/9334784  IRAC Mapping\\
                &              &             &               &  20597/15606016  MIPS Scan Map\\
                &              &             &               &  20597/15653632  MIPS Scan Map\\
                &              &             &               &  30570/20252928  IRAC Mapping\\
H 1617$-$155    & Sco X$-$1    & 16 19 55.07 &$-$15 38 24.80 &  20462/14727168  IRAC Mapping \\ 
                &              &             &               &  20462/14727424  MIPS Photometry\\
                &              &             &               &  20462/14727680  IRS Peakup Image\\
                &              &             &               &  50600/26536192  IRS Staring\\
4U 1624$-$49    &              & 16 28 02.83 &$-$49 11 54.61 &  191/9334528  IRAC Mapping\\
                &              &             &               &  20597/15610880  MIPS Scan Map\\
                &              &             &               &  20597/15620608  MIPS Scan Map\\ 
4U 1626$-$67    &              & 16 32 16.80 &$-$67 27 43.00 &  20224/14415360  IRAC Mapping\\
4U 1630$-$47    &              & 16 34 01.61 &$-$47 23 34.80 &  191/9326592  IRAC Mapping\\ 
                &              &             &               &  191/9330688  IRAC Mapping\\
                &              &             &               &  20597/15645184  MIPS Scan Map\\
                &              &             &               &  20597/20489984  MIPS Scan Map \\
                &              &             &               &  20597/15660800  MIPS Scan Map \\
                &              &             &               &  20597/20489728  MIPS Scan Map \\
4U 1636$-$536   &              & 16 40 55.50 &$-$53 45 05.00 &  No programs found\\ 
XTE J1650$-$500 &              & 16 50 00.98 &$-$49 57 43.60 &  No programs found\\ \hline
\end{tabular}
\end{table}

\setcounter{table}{0}
\begin{table}
\caption{continued}
\scriptsize
\begin{tabular}{lcrrl}
                &              &             &               & \\
{\bf Name\tablenotemark{a}} & {\bf Altern. Name}& {\bf RA (2000)} & {\bf DEC (2000)} & {\bf PID/AORkey, Instr. Mode, Notes} \\ \hline
                &              &             &               & \\
GRO J1655$-$40  &              & 16 54 00.14 &$-$39 50 44.90 &  3246/10524928  MIPS Photometry \\
                &              &             &               &  3246/10525184  MIPS Photometry \\
                &              &             &               &  3246/10525696  MIPS Photometry \\
                &              &             &               &  3246/10525440  MIPS Photometry  \\
                &              &             &               &  20301/14508800  IRAC Mapping \\
                &              &             &               &  20301/14509312  MIPS Photometry \\
                &              &             &               &  20669/14967040  IRAC Mapping \\
                &              &             &               &  30570/20220416  IRAC Mapping\\
                &              &             &               &  30570/21312768  IRAC Mapping \\
                &              &             &               &  30570/21311232  IRAC Mapping \\  
2A 1655+353     &  Her X$-$1   & 16 57 49.83 &   35 20 32.60 &  No programs found\\
XTE J1701$-$462 &              & 17 00 58.45 &$-$46 11 08.60 &  30570/21285376  IRAC Mapping\\
MXB 1659$-$298  &              & 17 02 06.50 &$-$29 56 44.10 &  No programs found\\
4U 1659$-$487   &  GX339$-$4   & 17 02 49.50 &$-$48 47 23.00 &  3246/10524416  MIPS Photometry\\
                &              &             &               &  3246/10524672  MIPS Photometry\\ 
3A 1702$-$363   &  GX349+2     & 17 05 44.50 &$-$36 25 23.00 &  30570/20246528  IRAC Mapping \\ 
                &              &             &               &  30570/21296384  IRAC Mapping\\
                &              &             &               &  30570/21324544  IRAC Mapping \\
                &              &             &               &  50670/26686976  MIPS Photometry\\
                &              &             &               &  50670/26687488  IRAC Mapping\\
4U 1702$-$429   &              & 17 06 15.31 &$-$43 02 08.69 &  No programs found\\
4U 1700+24      &              & 17 06 34.52 &   23 58 18.60 &  No programs found\\ 
4U 1705$-$250   &              & 17 08 14.60 &$-$25 05 29.00 &  No programs found\\ 
XTE J1709$-$267 &              & 17 09 30.40 &$-$26 39 19.90 &  No programs found\\ 
IGR J17098$-$3628&             & 17 09 45.93 &$-$36 27 58.20 &  30570/20246528  IRAC Mapping \\
                &              &             &               &  30570/21296384  IRAC Mapping \\
                &              &             &               &  30570/21324544  IRAC Mapping\\
GRO J1719$-$24  &              & 17 19 36.93 &$-$25 01 03.40 &  No programs found\\  
XTE J1720$-$318 &              & 17 19 58.99 &$-$31 45 01.25 &  50477/26273280 MIPS Scan Map\\  
IGR J17269$-$4737&             & 17 26 49.28 &$-$47 38 24.90 &  No programs found\\
4U 1728$-$16    &  GX 9+9      & 17 31 44.20 &$-$16 57 42.00 &  No programs found\\
4U 1728$-$34    &  GX 354$-$0  & 17 31 57.73 &$-$33 50 02.50 &  20201/14292992  IRAC Mapping \\
                &              &             &               &  20201/14293248  IRAC Mapping\\
                &              &             &               &  20201/14323200  IRAC Mapping\\
                &              &             &               &  20201/14322944  IRAC Mapping\\
3A 1728$-$247   &  GX 1+4      & 17 32 02.16 &$-$24 44 44.00 &  20119/14093056  IRAC Mapping\\
                &              &             &               &  20119/14099456  MIPS Scan Map\\
                &              &             &               &  20119/14099712  MIPS Scan Map\\
KS 1731$-$260   &              & 17 34 13.47 &$-$26 05 18.80 &  20119/14090496  IRAC Mapping\\
                &              &             &               &  20119/14099968  MIPS Scan Map\\
                &              &             &               &  20119/14100480  MIPS Scan Map\\
                &              &             &               &  30570/21277184  IRAC Mapping \\
                &              &             &               &  30570/21335552  IRAC Mapping \\
                &              &             &               &  30570/21291776  IRAC Mapping \\ 
                &              &             &               &  30570/21307136  IRAC Mapping\\
4U 1735$-$444   &              & 17 38 58.30 &$-$44 27 00.00 &  No programs found\\  
1H 1743$-$322   &              & 17 46 15.57 &$-$32 14 01.10 &  30570/20205824  IRAC Mapping \\
                &              &             &               &  30570/20210432  IRAC Mapping\\
                &              &             &               &  30570/21306624  IRAC Mapping\\
                &              &             &               &  30570/21313024  IRAC Mapping\\
                &              &             &               &  30594/20515840  MIPS\\
                &              &             &               &  30594/20517888  MIPS\\
                &              &             &               &  30594/20516864  MIPS \\
IGR J17473$-$2721&             & 17 47 18.08 &$-$27 20 38.70 &  20201/14309888  IRAC Mapping \\
                &              &             &               &  20201/14309632  IRAC Mapping \\
                &              &             &               &  20201/14339328  IRAC Mapping \\
                &              &             &               &  20201/14339584  IRAC Mapping\\
                &              &             &               &  20414/14658048  MIPS Scan Map  \\
                &              &             &               &  20414/14659072  MIPS Scan Map\\
1A 1744$-$361   &              & 17 48 13.15 &$-$36 07 57.02 &  No programs found\\
IGR J17497$-$2821&             & 17 49 38.04 &$-$28 21 17.37 &  3677/13368320  IRAC Mapping\\
                &              &             &               &  3677/13370624  IRAC Mapping\\
                &              &             &               &  20201/14308096  IRAC Mapping \\
                &              &             &               &  20201/14337792  IRAC Mapping \\
                &              &             &               &  20414/14657792  MIPS Scan Map \\
                &              &             &               &  20414/14658048  MIPS Scan Map\\
EXO 1747$-$214  &              & 17 50 24.52 &$-$21 25 19.90 &  30570/20208640  IRAC Mapping \\
4U 1755$-$33    &              & 17 58 40.00 &$-$33 48 27.00 &  No programs found\\
4U 1758$-$25    &  GX 5$-$1    & 18 01 08.22 &$-$25 04 42.46 &  20201/14314752  IRAC Mapping\\
                &              &             &               &  20201/14312704  IRAC Mapping\\
                &              &             &               &  20201/14342400  IRAC Mapping\\
                &              &             &               &  20201/14344448  IRAC Mapping\\
                &              &             &               &  30570/21293568  IRAC Mapping  \\ \hline
\end{tabular}
\end{table}

\setcounter{table}{0}
\begin{table}
\caption{continued}
\scriptsize
\begin{tabular}{lcrrl}
                &              &             &               & \\
{\bf Name\tablenotemark{a}} & {\bf Altern. Name}& {\bf RA (2000)} & {\bf DEC (2000)} & {\bf PID/AORkey, Instr. Mode, Notes} \\ \hline
                &              &             &               & \\
GRS 1758$-$258  &              & 18 01 12.40 &$-$25 44 36.10 &  20201/14312192  IRAC Mapping\\
                &              &             &               &  20201/14341888  IRAC Mapping \\
SAX J1808.4$-$3658&            & 18 08 27.60 &$-$36 58 43.90 &  No programs found\\
SAX J1810.8$-$2609&            & 18 10 44.47 &$-$26 09 01.20 &  No programs found\\
XTE J1814$-$338 &              & 18 13 39.04 &$-$33 46 22.30 &  No programs found\\
4U 1811$-$17    &  GX 13+1     & 18 14 31.55 &$-$17 09 26.70 &  146/12104192  IRAC Mapping \\
                &              &             &               &  146/12108800  IRAC Mapping \\
                &              &             &               &  20597/15598848  MIPS Scan Map\\
                &              &             &               &  20597/15634944  MIPS Scan Map\\
                &              &             &               &  20597/15632384  MIPS Scan Map\\
4U 1812$-$12    &              & 18 15 06.16 &$-$12 05 46.70 &  30570/21288448  IRAC Mapping\\
                &              &             &               &  30570/21309440  IRAC Mapping\\
4U 1813$-$14    &  GX 17+2     & 18 16 01.39 &$-$14 02 10.62 &  50670/26686720  MIPS Photometry\\
                &              &             &               &  50670/26687744  IRAC Mapping\\ 
AX J1824.5$-$2451&             & 18 24 30.00 &$-$24 51 00.00 &  40111/21745408  IRS Staring\\
4U 1823$-$00    &              & 18 25 22.02 &$-$00 00 43.00 &  20224/14415104  IRAC Mapping \\
2A 1822$-$371   &              & 18 25 46.80 &$-$37 06 19.00 &  1867/16040192  MIPS Scan Map \\
                &              &             &               &  1867/16040448  MIPS Scan Map\\
                &              &             &               &  1867/16040704  MIPS Scan Map\\
                &              &             &               &  1867/16040960  MIPS Scan Map\\
                &              &             &               &  1867/16041216  MIPS Scan Map \\
                &              &             &               &  1867/16041472  MIPS Scan Map\\
                &              &             &               &  1867/16041728  MIPS Scan Map\\
                &              &             &               &  1867/16041984  MIPS Scan Map\\
                &              &             &               &  1867/16042240  MIPS Scan Map\\
                &              &             &               &  1867/16042496  MIPS Scan Map\\
                &              &             &               &  1867/16042752  MIPS Scan Map\\
                &              &             &               &  1867/16043008  MIPS Scan Map\\
                &              &             &               &  1867/16043264  MIPS Scan Map\\
                &              &             &               &  1867/16043520  MIPS Scan Map\\
                &              &             &               &  1867/16043776  MIPS Scan Map\\
                &              &             &               &  1867/16044032  MIPS Scan Map\\
                &              &             &               &  1867/16044288  MIPS Scan Map \\
                &              &             &               &  1867/16044544  MIPS Scan Map\\
                &              &             &               &  1867/16044800  MIPS Scan Map\\
                &              &             &               &  1867/16045056  MIPS Scan Map\\
                &              &             &               &  1867/16057856  MIPS Scan Map\\
                &              &             &               &  1867/16058112  MIPS Scan Map\\
                &              &             &               &  1867/16058368  MIPS Scan Map \\
GS 1826$-$238   &              & 18 29 28.20 &$-$23 47 49.12 &  No programs found\\
2A 1837+049     &  Ser X$-$1   & 18 39 57.56 &   05 02 09.60 &  174/5716992  MIPS Scan Map\\
                &              &             &               &  BUT: position of LMXB not covered\\
XTE J1859+226   &              & 18 58 41.58 &   22 39 29.40 &  40948/23365120  IRAC Mapping\\
HETE J1900.1$-$2455&           & 19 00 08.65 &$-$24 55 13.70 &  No programs found\\
4U 1905+000     &              & 19 08 26.97 &   00 10 07.70 &  No programs found\\
XTE J1908+094   &              & 19 08 53.08 &   09 23 04.90 &  187/11963904  IRAC Mapping\\
                &              &             &               &  20597/15614720  MIPS Scan Map\\
4U 1908+005     &  Aql X$-$1   & 19 11 16.00 &   00 35 06.00 &  No programs found\\
GRS 1915+105    &              & 19 15 11.55 &   10 56 44.76 &  187/11970048  IRAC Mapping\\
                &              &             &               &  187/11973120  IRAC Mapping\\
                &              &             &               &  3352/10831616  IRAC Mapping\\
                &              &             &               &  3352/10832128  IRS Staring\\
                &              &             &               &  3352/10832384  IRS Staring\\
                &              &             &               &  3352/10832640  IRS Staring\\
                &              &             &               &  3352/10832896  IRS Staring\\
                &              &             &               &  3352/10831360  IRAC Mapping\\
                &              &             &               &  3352/10831104  IRAC Mapping\\
                &              &             &               &  3352/10831872  IRAC Mapping\\
                &              &             &               &  20232/14420480  IRAC Mapping\\
                &              &             &               &  20232/14420736  IRAC Mapping \\
                &              &             &               &  20232/14420992  IRAC Mapping\\
                &              &             &               &  20232/14420224  IRAC Mapping\\
                &              &             &               &  20232/14421248  IRS Staring \\
                &              &             &               &  20232/14422016  IRS Staring  \\
                &              &             &               &  20232/14421504  IRS Staring \\
                &              &             &               &  20232/14421760  IRS Staring\\
                &              &             &               &  20597/15599616  MIPS Scan Map \\
                &              &             &               &  20597/15614976  MIPS Scan Map  \\
4U 1916$-$05    &              & 19 18 47.87 &$-$05 14 17.09 &  20224/14414848  IRAC Mapping \\
3A 1954+319     &              & 19 55 42.33 &   32 05 49.10 &  No programs found\\
4U 1957+11      &              & 19 59 24.20 &   11 42 32.40 &  No programs found\\ \hline
\end{tabular}
\end{table}

\setcounter{table}{0}
\begin{table}
\caption{continued}
\scriptsize
\begin{tabular}{lcrrl}
                &              &             &               & \\
{\bf Name\tablenotemark{a}} & {\bf Altern. Name}& {\bf RA (2000)} & {\bf DEC (2000)} & {\bf PID/AORkey, Ins
tr. Mode, Notes} \\ \hline
                &              &             &               & \\
GS 2000+25      &              & 20 02 49.58 &   25 14 11.30 &  40948/23365632  IRAC Mapping\\
XTE J2012+381   &              & 20 12 37.71 &   38 11 01.10 &  20726/15055616  MIPS Photometry\\
                &              &             &               &  20726/15055872  IRAC Mapping\\
GS 2023+338     &              & 20 24 03.80 &   33 52 03.20 &  3289/10703104  MIPS Photometry \\
                &              &             &               &  3289/10704128  IRAC Mapping \\
XTE J2123$-$058 &              & 21 23 14.54 &$-$05 47 53.20 &  No programs found\\
4U 2129+47      &              & 21 31 26.20 &   47 17 24.00 &  No programs found\\
4U 2142+38      &  Cyg X$-$2   & 21 44 41.20 &   38 19 18.00 &  50670/26687232  MIPS Photometry \\
                &              &             &               &  50670/26688000  IRAC Mapping\\
\tablenotetext{a}{nomenclature from Liu et al. (2007)}
\end{tabular}
\end{table}

\vspace{-0.5cm}

\setcounter{table}{1}
\begin{table}[t!]
\caption{Results}
\scriptsize
\begin{tabular}{lcccccl}
           &   & & & & & \\
{\bf Name} & {\bf F3.6}& {\bf F4.5} & {\bf F5.8} & {\bf F8.0} & {\bf F24} &
{\bf Notes} \\ 
          & (\bf mJy)  &(\bf mJy) &(\bf mJy) & (\bf mJy)&(\bf mJy) & \\ \hline 
           &   & & & & & \\
LMC X$-$2     & 0.038    & yes(f) & no      & no      & no       & SAGE Winter'07 catalog, this work\\
X0614+091     & 0.16     & 0.18   & 0.2     & 0.25    & yes      & Migliari et al.\ 2006; this work (24$\mu$m) \\
X0620$-$003   & ...  & 0.4    & ... & 0.29    & 0.14     & Gallo et al.\ 2007 \\
              & ...  & 0.45   & ... & 0.25    & 0.05     & Muno et al.\ 2006 \\
X0918$-$549   & no       & no     & no      & no      & ...  & this work, north component of pair detected\\
J1118+480     & ...  & 0.069  & ... & 0.059   & $< 0.05$ & Gallo et al.\ 2007   \\
              & ...  & 0.046  & ... & 0.045   & $< 0.02$ & Muno et al.\ 2006  \\
X1323$-$62    & no       & no     & no      & no      & no       & Counterpart uncertain \\
Cen X$-$4     & ...  & 0.20   & ... & 0.095   & $< 0.03$ & Muno et al.\ 2006 \\
Cir X$-$1     & 18.55    & 15.26  & 12.40   & 8.75    &   6      & GLIMPSE Spring'07 catalog, this work (24$\mu$m) \\
X1543$-$62    & no       & no     & no      & no      & ...  & this work\\
X1608$-$52    & yes(f)   & yes(f) & no      & no      & no       & this work, very crowded\\
Sco X$-$1     &  10      &   7    &  4      & 1.5     & 1        & Wachter et al.\ 2006 \\
X1624$-$490   & no       & no     & no      & no      & no       & this work, very crowded\\
X1626$-$67    & yes(f)   & yes(f) & yes(f)  & no      & ...  & this work \\
X1630$-$47    & no       & no     & no      & no      & no       & this work, too crowded \\
X1655$-$40    & 5.67     & 3.90   & 3.02    & 2.40    & $<0.54 - 2.07$& Migliari et al.\ 2007\\
GX 339$-$4    & ...  & ...& ... & ... & 0.135    & Tomsick et al. 2004\\
XJ1701$-$462  & yes      & yes    & yes     & yes     & ...  & this work, correct ctpt? \\
GX 349+2      & yes      & yes    & yes     & yes(f)  & ...  & this work\\
IGR J17098$-$3628& no    &  no    & no      & no      & ...  & this work, too crowded\\
GX 354$-$0    & yes(f)   & yes(f) & yes(f)  & no:     & ...  & this work\\
GX 1+4        & yes(s)   & yes    & yes     & yes     & yes      & this work, saturated ch1\\ 
X1731$-$260   & no       & no     & no      & no      & no       & this work, close neighbor detected\\  
X1743$-$322   & yes(f)   & yes(f) & yes(f)  & no      & no       & this work, blend of three sources, too crowded\\ 
IGR J1747$-$2721& no     & no     & no      & no      & no       & this work, too crowded\\
IGR J17497$-$2821& yes(f):& yes(f):& yes(f):& no      & no       & this work, too crowded\\ 
X1747$-$214   & ...  & no     & ... & no      & ...  & this work, ctpt uncertain, blend of 3 sources\\
GX 5$-$1      & yes      & yes    & no:     & no      & ...  & this work, close blend of two sources\\
X1758$-$258   & yes      & yes    & yes     & no      & ...  & this work \\
GX 13+1       & 7.82     & 6.13   & 4.20    & 2.49    & no       & GLIMPSE Spring'07 catalog, this work (24$\mu$m)\\
X1812$-$12    & yes(f):  & yes(f):& yes(f)  & no      & ...  & this work \\
X1822$-$00    & yes(f)   & yes(f) & yes(f)  & no      & ...  & this work \\
XTE J1908+094 & no       & no     & no      & no      & no       & this work, too crowded \\
X1915+105     & 4.66     & 4.96   & 4.63    & 2.96    & ~28.3    & GLIMPSE Spring'07 catalog, this work (24$\mu$m)\\
X1916$-$05    & yes      & yes    & yes(f)  & yes     & ...  & this work, blended with neighbor \\
XTE J2012+381 & yes      & yes    & no      & no      & no       & this work, blended with neighbor \\
V404 Cyg      & ...  & 3.3    & ... & 1.8     & 0.4      & Gallo et al.\ 2007 \\
              & ...  & 3.0    & ... & 1.45    & 1.53     & Muno et al.\ 2006\\
\end{tabular}
\end{table}

\end{document}